\documentclass[twocolumn,showpacs,preprintnumbers]{revtex4}%
\usepackage{amssymb}
\usepackage{amsmath}
\usepackage{graphicx}
\usepackage{dcolumn}
\usepackage{bm}
\usepackage{amsfonts}%
\begin{document}
\title{Phase Space Distribution Near Self-Excited Oscillation Threshold}
\author{Yuvaraj Dhayalan}
\affiliation{Department of Electrical Engineering, Technion, Haifa 32000 Israel}
\author{Ilya Baskin}
\affiliation{Department of Electrical Engineering, Technion, Haifa 32000 Israel}
\author{Keren Shlomi}
\affiliation{Department of Electrical Engineering, Technion, Haifa 32000 Israel}
\author{Eyal Buks}
\affiliation{Department of Electrical Engineering, Technion, Haifa 32000 Israel}
\date{\today }

\begin{abstract}
We study phase space distribution of an optomechanical cavity near the
threshold of self-excited oscillation. A fully on-fiber optomechanical cavity
is fabricated by patterning a suspended metallic mirror on the tip of the
fiber. Optically induced self-excited oscillation of the suspended mirror is
observed above a threshold value of the injected laser power. A theoretical
analysis based on Fokker-Planck equation evaluates the expected phase space
distribution near threshold. A tomography technique is employed for extracting
phase space distribution from the measured reflected optical power vs. time in
steady state. Comparison between theory and experimental results allows the
extraction of the device parameters.

\end{abstract}
\maketitle



%It is always \today, today,
%but any date may be explicitly specified

%PACS, the Physics and Astronomy
%Classification Scheme.

Optomechanical cavities are currently a subject of intense basic and applied
study \cite{Braginsky&Manukin_67,
Hane_179,Gigan_67,Metzger_1002,Kippenberg_1172,Favero_104101,Marquardt2009}.
Optomechanical cavities can be employed in various sensing \cite{Rugar1989,
Arcizet2006a, Forstner2012,Weig2013} and photonics applications
\cite{Lyshevski&Lyshevski_03,Stokes_et_al_90,
Hossein-Zadeh_Vahala_10,Wu_et_al_06,
MattEichenfield2007,Bahl2011,Flowers-Jacobs2012}. Moreover, such systems may
allow experimental study of the crossover from classical to quantum mechanics
\cite{Thompson_72, Meystre2013,Kimble_et_al_01, Carmon_et_al_05,
Arcizet_et_al_06, Gigan_67, Jayich_et_al_08, Schliesser_et_al_08,
Genes_et_al_08, Teufel_et_al_10} (see Ref. \cite{Poot_273} for a recent
review). When the finesse of the optical cavity that is employed for
constructing the optomechanical cavity is sufficiently high, the coupling to
the mechanical resonator that serves as a vibrating mirror is typically
dominated by the effect of radiation pressure
\cite{Kippenberg_et_al_05,Rokhsari2005,
Arcizet2006,Gigan_et_al_06,Cooling_Kleckner06, Kippenberg_1172}. On the other
hand, bolometric effects can contribute to the optomechanical coupling when
optical absorption by the vibrating mirror is significant \cite{Metzger_1002,
Jourdan_et_al_08,Marino&Marin2011PRE, Metzger_133903, Restrepo_860,
Liberato_et_al_10,Marquardt_103901, Paternostro_et_al_06,Yuvaraj_430}. In
general, bolometric effects play an important role in relatively large
mirrors, in which the thermal relaxation rate is comparable to the mechanical
resonance frequency \cite{Aubin_et_al_04, Marquardt_103901,
Paternostro_et_al_06, Liberato_et_al_10_PRA}. Phenomena such as mode cooling
and self-excited oscillation
\cite{Hane_179,Kim_1454225,Aubin_1018,Carmon_223902,Marquardt_103901,Corbitt_021802,Carmon_123901,Metzger_133903}
have been shown in systems in which bolometric effects are dominant
\cite{Metzger_133903, Metzger_1002, Aubin_et_al_04,
Jourdan_et_al_08,Zaitsev_046605,Zaitsev_1589}.%

%TCIMACRO{\FRAME{ftbpFU}{3.2396in}{2.8643in}{0pt}{\Qcb{A schematic drawing of
%sample A and the experimental set-up. An on-fiber optomechanical cavity is
%excited by a laser. The reflected light intensity is measured and analyzed.
%The inset shows a typical trace of the photodetector voltage vs. time measured
%by the oscilloscope above self-excited oscillation threshold with $\Delta
%P_{\mathrm{L}}/P_{\mathrm{LC}}=0.15$.}}{\Qlb{Fig setup}}{fig1.eps}%
%maintain-aspect-ratio TRUE;  display "ICON";  valid_file "F";
%width 3.2396in;  height 2.8643in;  depth 0pt;  original-width 8.3636in;
%original-height 7.3864in;  cropleft "0";  croptop "1";  cropright "1";
%cropbottom "0";  filename 'Fig1.eps';file-properties "XNPEU";}} }%
%BeginExpansion
\begin{figure}
[ptb]
\begin{center}
\includegraphics[
height=2.8643in,
width=3.2396in
]%
{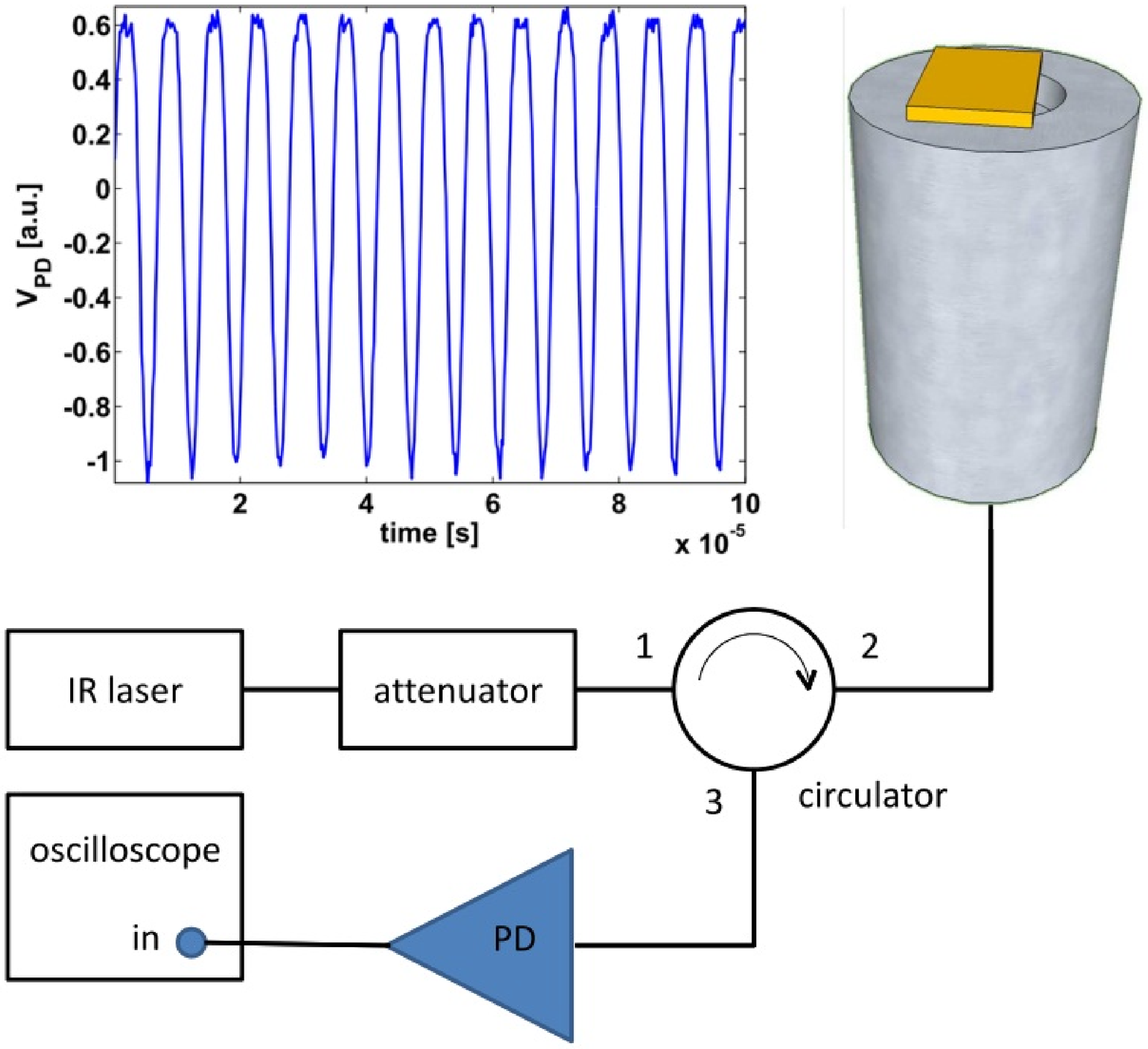}%
\caption{A schematic drawing of sample A and the experimental set-up. An
on-fiber optomechanical cavity is excited by a laser. The reflected light
intensity is measured and analyzed. The inset shows a typical trace of the
photodetector voltage vs. time measured by the oscilloscope above self-excited
oscillation threshold with $\Delta P_{\mathrm{L}}/P_{\mathrm{LC}}=0.15$.}%
\label{Fig setup}%
\end{center}
\end{figure}
%EndExpansion

Recently, it has been demonstrated that optomechanical cavities can be
fabricated on the tip of an optical fiber \cite{iannuzzi06, Ma2010,
iannuzzi10, iannuzzi11, Jung2011,
Butsch2012,Albri2013,Shkarin_1306_0613,Baskin_1210_7327}. These miniature
devices appear to be very promising for sensing applications. However, their
operation requires external driving of the on-fiber mechanical resonator.
Traditional driving using either electrical or magnetic actuation, however, is
hard to implement with a mechanical resonator on the tip of an optical fiber;
a limitation that can be overcome by optical actuation schemes
\cite{Iannuzzi2013}.

\begin{figure}
[ptb]
\begin{center}
\includegraphics[
height=2.6524in,
width=3.2396in
]%
{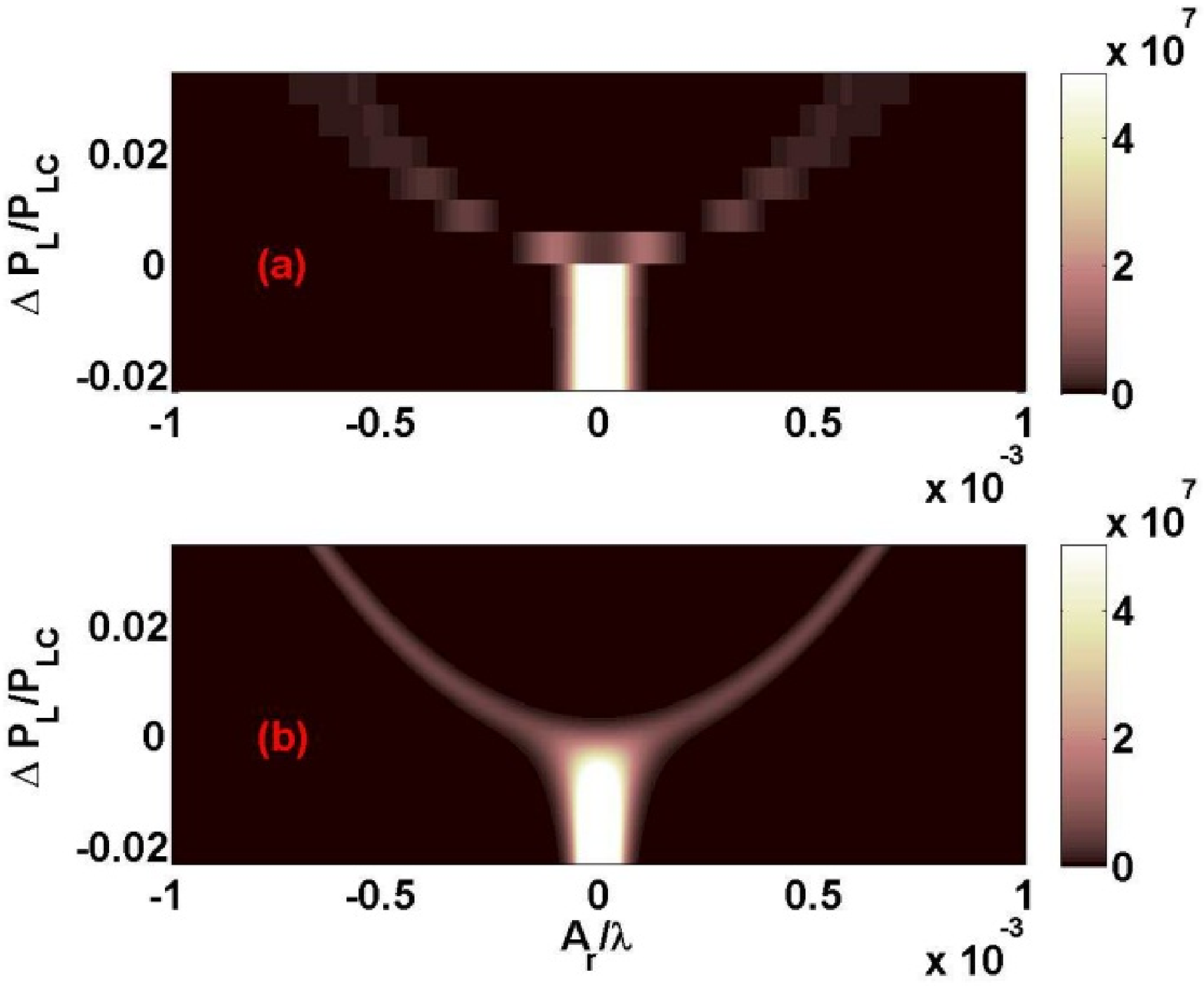}%
\caption{The dependence on laser power $P_{\mathrm{L}}$ - sample A. (a) Phase
space distribution extracted from the measured probability distribution
function $w\left(  X_{\phi}^{\prime}\right)  $ using Eq. (\ref{P(A_x,A_y)}).
(b) Phase space distribution calculated using Eq. (\ref{Wig steady state}).
The following device parameters have been employed for generating the plot in
panel (b): $\omega_{0}=2\pi\times144\operatorname{kHz}$, $\lambda
=1.55\operatorname{\mu m}$, $T_{\mathrm{eff}}=300\operatorname{K}$,
$s_{\mathrm{D}}=0.8$ and $\gamma_{2}\lambda^{2}/\gamma_{0}=8\times10^{4}$.}%
\label{Fig Wigner vs P_L}%
\end{center}
\end{figure}

\begin{figure}
[ptb]
\begin{center}
\includegraphics[
height=2.6852in,
width=3.2396in
]%
{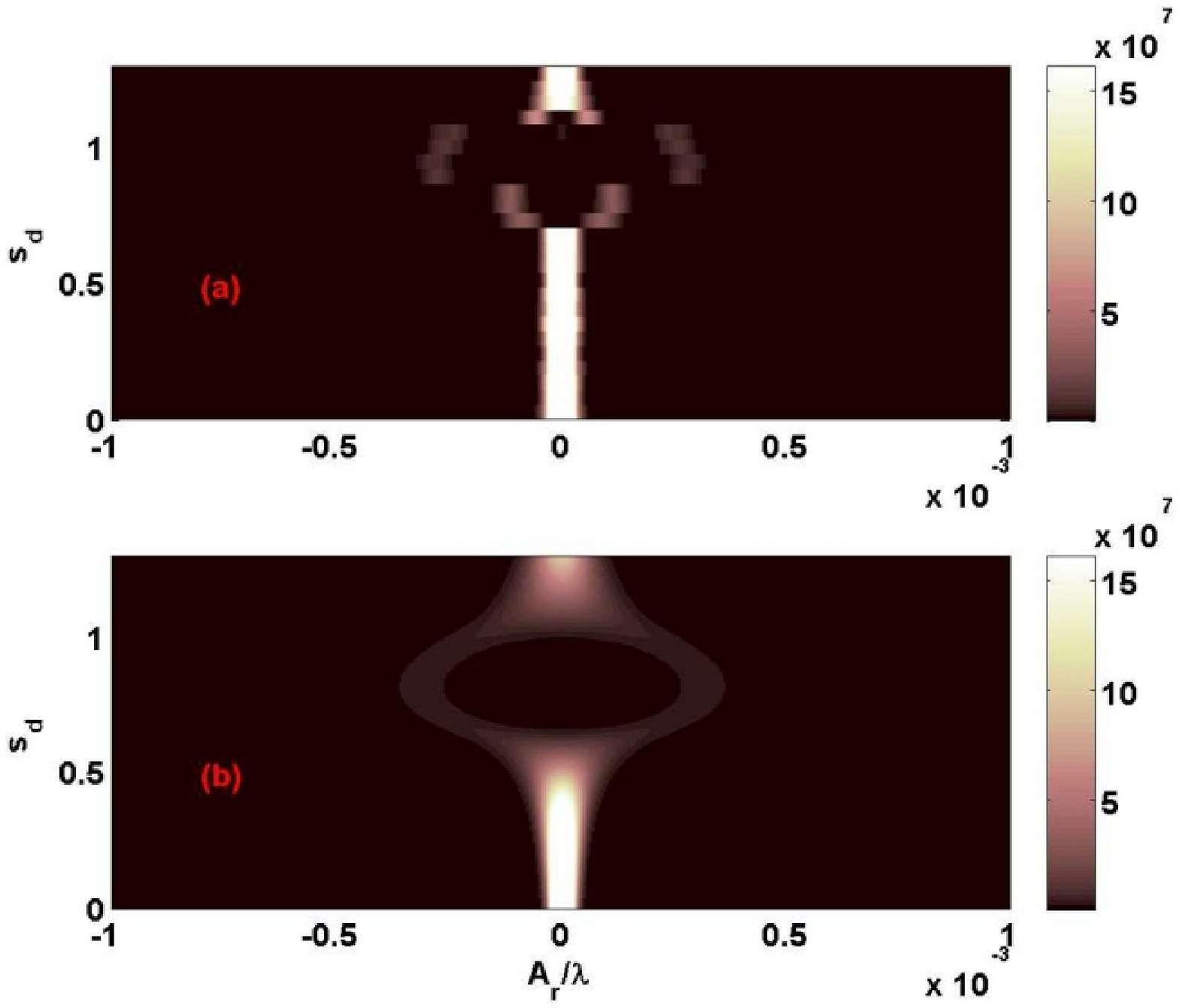}%
\caption{The dependence on wavelength $\lambda$ - sample B. The static mirror
of the optomechanical cavity is provided by a fiber Bragg grating (FBG) mirror
(made using a standard phase mask technique \cite{Anderson_566}, grating
period of $0.527\operatorname{\mu m}$ and length $\approx8\operatorname{mm}$)
with the reflectivity band of $0.4\operatorname{nm}$ full width at half
maximum (FWHM) centered at $1550\operatorname{nm}$. The length of the optical
cavity was $10\operatorname{mm}$. (a) Phase space distribution extracted from
the measured probability distribution function $w\left(  X_{\phi}^{\prime
}\right)  $ using Eq. (\ref{P(A_x,A_y)}). The laser wavelength is varied from
the cavity resonance value of $\lambda_{\mathrm{R}}%
=1.5451702\operatorname{\mu m}$, for which the detuning factor $s_{\mathrm{D}%
}$ vanishes, to $1.5451942\operatorname{\mu m}$, for which $s_{\mathrm{D}%
}=1.2986$. (b) Phase space distribution calculated using Eq.
(\ref{Wig steady state}). The following device parameters have been employed
for generating the plot in panel (b): $\omega_{0}=2\pi\times
225\operatorname{kHz}$, $m=1.1\times10^{-12}\operatorname{kg}$, $\beta
_{+}=0.68$, $T_{\mathrm{eff}}=300\operatorname{K}$ and $\gamma_{2}\lambda
^{2}/\gamma_{0}=5\times10^{5}$.}%
\label{Fig Wigner vs WL}%
\end{center}
\end{figure}

In this paper we study a configuration of an on-fiber optomechanical cavity
and demonstrate that self-excited oscillation can be optically induced by
injecting a monochromatic laser light into the fiber. The optomechanical
cavity is formed between the vibrating mirror that is fabricated on the tip of
a single mode optical fiber and an additional static reflector. The results
seen in Figs. \ref{Fig setup} and \ref{Fig Wigner vs P_L} below have been
obtained with a sample (labeled as sample A) in which the static reflector is
the glass-vacuum interface at the fiber's tip, whereas the results seen in
Fig. \ref{Fig Wigner vs WL} have been obtained with a sample (labeled as
sample B) in which the static reflector is a fiber Bragg grating (FBG). For
both samples, optically-induced self-excited oscillation is attributed to the
bolometric optomechanical coupling between the optical mode and the mechanical
resonator \cite{Zaitsev_046605,Zaitsev_1589}.

Optomechanical cavities operating in the region of self-excited oscillation
can be employed for sensing applications. Such a device can sense physical
parameters that affect the mechanical properties of the suspended mirror (e.g.
absorbed mass, heating by external radiation, acceleration, etc.). The
sensitivity of such a sensor is limited by the phase noise of the self-excited
oscillation. Here we experimentally measure the phase space distribution of
the mechanical element near the threshold of self-excited oscillation and
compare the results with theoretical predictions.

The optomechanical cavity schematically shown in Fig. \ref{Fig setup} was
fabricated on the flat polished tip of a single mode fused silica optical
fiber having outer diameter of $126%
%TCIMACRO{\unit{\U{3bc}m}}%
%BeginExpansion
\operatorname{\mu m}%
%EndExpansion
$ (Corning SMF-28 operating at wavelength band around $\lambda=1550%
%TCIMACRO{\unit{nm}}%
%BeginExpansion
\operatorname{nm}%
%EndExpansion
$) held in a zirconia ferrule. Thermal evaporation through a mechanical mask
was employed for patterning a metallic rectangle (see Fig. \ref{Fig setup})
made of a $10%
%TCIMACRO{\unit{nm}}%
%BeginExpansion
\operatorname{nm}%
%EndExpansion
$ thick chromium layer and a $200%
%TCIMACRO{\unit{nm}}%
%BeginExpansion
\operatorname{nm}%
%EndExpansion
$ thick gold layer. The metallic rectangle, which serves as a mirror, covers
almost the entire fiber cross section. However, a small segment is left open
in order to allow suspension of the mirror, which was done by etching
approximately $12%
%TCIMACRO{\unit{\U{3bc}m}}%
%BeginExpansion
\operatorname{\mu m}%
%EndExpansion
$ of the underlying silica in 7\% HF acid ($90%
%TCIMACRO{\unit{min}}%
%BeginExpansion
\operatorname{min}%
%EndExpansion
$ etch time at room temperature). The suspended mirror remained supported by
the zirconia ferrule, which is resistant to HF.

Monochromatic light was injected into the fiber of sample A from a laser
source having wavelength $\lambda=1550.08%
%TCIMACRO{\unit{nm}}%
%BeginExpansion
\operatorname{nm}%
%EndExpansion
$ and an adjustable power level $P_{\mathrm{L}}$. The laser was connected
through an optical circulator, that allowed the measurement of the reflected
light intensity ($P_{\mathrm{R}}$) by a fast responding photodetector. The
detected signal was analyzed by an oscilloscope and a spectrum analyzer (see
Fig. \ref{Fig setup}). The experiments were performed in vacuum (at residual
pressure below $0.01%
%TCIMACRO{\unit{Pa}}%
%BeginExpansion
\operatorname{Pa}%
%EndExpansion
$). The angular frequency of the fundamental mode of the suspended mirror
$\omega_{0}=2\pi\times144%
%TCIMACRO{\unit{kHz}}%
%BeginExpansion
\operatorname{kHz}%
%EndExpansion
$ was estimated by the frequency of thermal oscillation measured at low input
laser power. When the injected laser power $P_{\mathrm{L}}$ exceeds a
threshold value given by $P_{\mathrm{LC}}=4.3%
%TCIMACRO{\unit{mW}}%
%BeginExpansion
\operatorname{mW}%
%EndExpansion
$, optically-induced self-excited oscillation of the vibrating mirror is
observed (see Fig. \ref{Fig setup}).

In the limit of small displacement the dynamics of the system can be
approximately described using a single evolution equation \cite{Zaitsev_1589}.
The theoretical model that is used to derive the evolution equation is briefly
described below. Note that some optomechanical effects that where taken into
account in the theoretical modeling \cite{Zaitsev_1589} were found
experimentally to have a negligible effect on the dynamics
\cite{Zaitsev_046605} (e.g. the effect of radiation pressure). In what follows
such effects are disregarded.%

%TCIMACRO{\FRAME{ftbpFU}{3.2396in}{2.6524in}{0pt}{\Qcb{The dependence on laser
%power $P_{\mathrm{L}}$ - sample A. (a) Phase space distribution extracted from
%the measured probability distribution function $w\left(  X_{\phi}^{\prime
%}\right)  $ using Eq. (\ref{P(A_x,A_y)}). (b) Phase space distribution
%calculated using Eq. (\ref{Wig steady state}). The following device parameters
%have been employed for generating the plot in panel (b): $\omega_{0}%
%=2\pi\times144\unit{kHz}$, $\lambda=1.55\unit{\U{3bc}m}$, $T_{\mathrm{eff}%
%}=300\unit{K}$, $s_{\mathrm{D}}=0.8$ and $\gamma_{2}\lambda^{2}/\gamma
%_{0}=8\times10^{4}$.}}{\Qlb{Fig Wigner vs P_L}}{fig2.eps}%
%maintain-aspect-ratio TRUE;  display "ICON";  valid_file "F";
%width 3.2396in;  height 2.6524in;  depth 0pt;  original-width 7.4158in;
%original-height 6.0623in;  cropleft "0";  croptop "1";  cropright "1";
%cropbottom "0";  filename 'Fig2.eps';file-properties "XNPEU";}} }%
%BeginExpansion
%EndExpansion

The micromechanical mirror in the optical cavity is treated as a mechanical
resonator with a single degree of freedom $x$ having mass $m$ and linear
damping rate $\gamma_{0}$ (when it is decoupled from the optical cavity). It
is assumed that the angular resonance frequency $\omega_{\mathrm{m}}$ of the
mechanical resonator depends on the temperature $T$ of the suspended mirror.
For small deviation of $T$ from the base temperature $T_{0}$ (i.e. the
temperature of the supporting substrate) $\omega_{\mathrm{m}}$ is taken to be
given by $\omega_{\mathrm{m}}=\omega_{0}-\beta\left(  T-T_{0}\right)  $, where
$\beta$ is a constant. Furthermore, to model the effect of thermal deformation
\cite{Metzger_133903} it is assumed that a temperature dependent force given
by $F_{\text{\textrm{th}}}=\theta\left(  T-T_{0}\right)  $, where $\theta$ is
a constant, acts on the mechanical resonator \cite{Yuvaraj_430}.

The intra-cavity optical power incident on the suspended mirror, which is
denoted by $P_{\mathrm{L}}I\left(  x\right)  $, where $P_{\mathrm{L}}$ is the
injected laser power, depends on the mechanical displacement $x$ (i.e. on the
length of the optical cavity). For small $x$, the expansion $I\left(
x\right)  \simeq I_{0}+I_{0}^{\prime}x+\left(  1/2\right)  I_{0}^{\prime
\prime}x^{2}$ is employed, where a prime denotes differentiation with respect
to the displacement $x$. The time evolution of the effective temperature $T$
is governed by the thermal balance equation $\dot{T}=\kappa\left(
T_{0}-T\right)  +\eta P_{\mathrm{L}}I\left(  x\right)  $, where overdot
denotes differentiation with respect to time $t$, $\eta$ is the heating
coefficient due to optical absorption and $\kappa$ is the thermal rate.

The function $I\left(  x\right)  $ depends on the properties of the optical
cavity formed between the suspended mechanical mirror and the on-fiber static
reflector (the glass-vacuum interface on the fiber's tip for sample A or FBG
for sample B). The finesse of the optical cavity is limited by loss mechanisms
that give rise to optical energy leaking out of the cavity. The main escape
routes are through the on-fiber static reflector, through absorption by the
metallic mirror, and through radiation. The corresponding transmission
probabilities are respectively denoted by $T_{\mathrm{B}}$, $T_{\mathrm{A}}$
and $T_{\mathrm{R}}$. In terms of these parameters the function $I\left(
x\right)  $ is given by \cite{Zaitsev_046605}%
\begin{equation}
I\left(  x\right)  =\frac{\beta_{\mathrm{F}}\left(  1-\frac{\beta_{-}^{2}%
}{\beta_{+}^{2}}\right)  \beta_{+}^{2}}{1-\cos\frac{4\pi x_{\mathrm{D}}%
}{\lambda}+\beta_{+}^{2}}\;,\label{I(x_D)}%
\end{equation}
where $x_{\mathrm{D}}=x-x_{\mathrm{R}}$ is the displacement of the mirror
relative to a point $x_{\mathrm{R}}$, at which the energy stored in the
optical cavity in steady state obtains a local maximum, $\beta_{\pm}%
^{2}=\left(  T_{\mathrm{B}}\pm T_{\mathrm{A}}\pm T_{\mathrm{R}}\right)
^{2}/8$ and where $\beta_{\mathrm{F}}$ is the cavity finesse, which is related
to $\beta_{+}$ by $\beta_{\mathrm{F}}=\omega_{\mathrm{FSR}}/\omega
_{\mathrm{C}}\beta_{+}$, where $\omega_{\mathrm{FSR}}$ is the free spectral
range and $\omega_{\mathrm{C}}$ is the angular cavity resonance frequency. The
reflection probability $R_{\mathrm{C}}=P_{\mathrm{R}}/P_{\mathrm{L}}$ is given
in steady state by \cite{Yurke_5054,Zaitsev_046605} $R_{\mathrm{C}}=1-I\left(
x\right)  /\beta_{\mathrm{F}}$.

The displacement $x\left(  t\right)  $ can be expressed in terms of the
complex amplitude $A$ as $x\left(  t\right)  =x_{0}+2\operatorname{Re}A$,
where $x_{0}$, which is given by $x_{0}=\eta\theta P_{\mathrm{L}}I_{0}%
/\kappa\omega_{0}^{2}$, is the optically-induced static displacement. For a
small displacement, the evolution equation for the complex amplitude $A$ is
found to be given by \cite{Zaitsev_1589}%
\begin{equation}
\dot{A}+\left(  \Gamma_{\mathrm{eff}}+i\Omega_{\mathrm{eff}}\right)
A=\xi\left(  t\right)  \;, \label{A dot}%
\end{equation}
where both the effective resonance frequency $\Omega_{\mathrm{eff}}$ and the
effective damping rate $\Gamma_{\mathrm{eff}}$ are real even functions of
$\left\vert A\right\vert $. To second order in $\left\vert A\right\vert $ they
are given by%
\begin{equation}
\Gamma_{\mathrm{eff}}=\Gamma_{0}+\Gamma_{2}\left\vert A\right\vert
^{2}\;,\;\Omega_{\mathrm{eff}}=\Omega_{0}+\Omega_{2}\left\vert A\right\vert
^{2}\;, \label{Gamma_eff,Omega_eff}%
\end{equation}
where $\Gamma_{0}=\gamma_{0}+\eta\theta P_{\mathrm{L}}I_{0}^{\prime}%
/2\omega_{0}^{2}$, $\Gamma_{2}=\gamma_{2}+\eta\beta P_{\mathrm{L}}%
I_{0}^{\prime\prime}/4\omega_{0}$, $\gamma_{2}$ is the mechanical nonlinear
quadratic damping rate \cite{Zaitsev_859}, $\Omega_{0}=\omega_{0}-\eta\beta
P_{\mathrm{L}}I_{0}/\kappa$ and $\Omega_{2}=-\eta\beta P_{\mathrm{L}}%
I_{0}^{\prime\prime}/\kappa$. Note that the above expressions for
$\Gamma_{\mathrm{eff}}$ and $\Omega_{\mathrm{eff}}$ are obtained by making the
following assumptions: $\beta x_{0}\ll\theta/2\omega_{0}$, $\theta\kappa
^{2}\ll\beta\omega_{0}^{3}\lambda$, where $\lambda$ is the optical wavelength,
and $\kappa\ll\omega_{0}$, all of which typically hold experimentally
\cite{Zaitsev_046605}. The fluctuating term \cite{Risken_Fokker-Planck}
$\xi\left(  t\right)  =\xi_{x}\left(  t\right)  +i\xi_{y}\left(  t\right)  $,
where both $\xi_{x}$ and $\xi_{y}$ are real, represents white noise and the
following is assumed to hold: $\left\langle \xi_{x}\left(  t\right)  \xi
_{x}\left(  t^{\prime}\right)  \right\rangle =\left\langle \xi_{y}\left(
t\right)  \xi_{y}\left(  t^{\prime}\right)  \right\rangle =2\Theta
\delta\left(  t-t^{\prime}\right)  $ and $\left\langle \xi_{x}\left(
t\right)  \xi_{y}\left(  t^{\prime}\right)  \right\rangle =0$, where
$\Theta=\gamma_{0}k_{\mathrm{B}}T_{\mathrm{eff}}/4m\omega_{0}^{2}$,
$k_{\mathrm{B}}$ is the Boltzmann's constant and $T_{\mathrm{eff}}$ is the
effective noise temperature. In cylindrical coordinates, $A$ is expressed as
$A=A_{r}e^{iA_{\theta}}$, where $A_{r}=\left\vert A_{r}\right\vert $ and
$A_{\theta}$ is real \cite{Hempstead_350}. The Langevin equation for the
radial coordinate $A_{r}$ can be written as%
\begin{equation}
\dot{A}_{r}+\frac{\partial\mathcal{H}}{\partial A_{r}}=\xi_{r}\left(
t\right)  \;, \label{A_r dot H}%
\end{equation}
where $\mathcal{H}\left(  A_{r}\right)  =\Gamma_{0}A_{r}^{2}/2+\Gamma_{2}%
A_{r}^{4}/4$ and the white noise term $\xi_{r}\left(  t\right)  $ satisfies
$\left\langle \xi_{r}\left(  t\right)  \xi_{r}\left(  t^{\prime}\right)
\right\rangle =2\Theta\delta\left(  t-t^{\prime}\right)  $.%

%EndExpansion

Consider the case where $\Gamma_{2}>0$, for which a supercritical Hopf
bifurcation occurs when the linear damping coefficient $\Gamma_{0}$ vanishes.
Above threshold, i.e. when $\Gamma_{0}$ becomes negative, Eq. (\ref{A_r dot H}%
) has a steady state solution (when noise is disregarded) at the point
$r_{0}=\sqrt{-\Gamma_{0}/\Gamma_{2}}$ [see Eq. (\ref{Gamma_eff,Omega_eff})].
The Langevin equation (\ref{A_r dot H}) yields a corresponding Fokker-Planck
equation, which in turn can be used to evaluate the normalized phase space
probability distribution function in steady state
\cite{Hempstead_350,Risken_Fokker-Planck}, which is found to be given by%
\begin{equation}
\mathcal{P}=\frac{e^{-\left(  \frac{A_{r}}{\delta_{0}}\right)  ^{2}-\frac
{1}{4\nu^{2}}\left(  \frac{A_{r}}{\delta_{0}}\right)  ^{4}}}{\pi^{\frac{3}{2}%
}\delta_{0}^{2}\nu e^{\nu^{2}}\left(  1-\operatorname{erf}\nu\right)  }\;,\;
\label{Wig steady state}%
\end{equation}
where $\delta_{0}^{2}=2\Theta/\Gamma_{0}$ and where $\nu=\Gamma_{0}%
/\sqrt{4\Gamma_{2}\Theta}$. Note that $\mathcal{P}$ is independent on the
angle $A_{\theta}$.

Experimentally, the technique of state tomography can be employed for
extracting phase space probability distribution from measured displacement of
the mechanical resonator. The normalized homodyne observable $X_{\phi}$ with a
real phase $\phi$ is defined by $X_{\phi}=2^{-1/2}\left(  A^{\ast}e^{i\phi
}+Ae^{-i\phi}\right)  $. Let $w\left(  X_{\phi}^{\prime}\right)  $ be the
normalized probability distribution function of the observable $X_{\phi}$. In
general, with the help of the inverse Radon transform, the phase space
probability distribution function $\mathcal{P}$ can be expressed in terms of
the probability distribution functions $w\left(  X_{\phi}^{\prime}\right)  $
\cite{Vogel_2847}. With a CW laser excitation, in steady state, $w\left(
X_{\phi}^{\prime}\right)  $ is expected to be $\phi$ independent. For such a
case one finds that%
\begin{equation}
\mathcal{P}=\frac{1}{2\pi}\int\limits_{0}^{\infty}\mathrm{d}\zeta\;\tilde
{w}\left(  \zeta\right)  \zeta J_{0}\left(  \zeta\sqrt{A_{x}^{2}+A_{y}^{2}%
}\right)  \;, \label{P(A_x,A_y)}%
\end{equation}
where the notation $J_{n}$ is used to label Bessel functions of the first
kind, and where $\tilde{w}\left(  \zeta\right)  $, which is given by
$\tilde{w}\left(  \zeta\right)  =%
%TCIMACRO{\dint \nolimits_{-\infty}^{\infty}}%
%BeginExpansion
{\displaystyle\int\nolimits_{-\infty}^{\infty}}
%EndExpansion
\mathrm{d}X_{\phi}^{\prime}\;w\left(  X_{\phi}^{\prime}\right)  e^{-i\zeta
X_{\phi}^{\prime}}$, is the characteristic function of $w\left(  X_{\phi
}^{\prime}\right)  $, i.e. $\mathcal{P}$ is found to be the the Hankel
transform of the characteristic function $\tilde{w}\left(  \zeta\right)  $.

Sample A, which is seen schematically in Fig. \ref{Fig setup}, was used to
study the dependence of phase space distribution on laser power. To that end,
the photodetector signal (see Fig. \ref{Fig setup}) was recorded over a time
period of $2%
%TCIMACRO{\unit{ms}}%
%BeginExpansion
\operatorname{ms}%
%EndExpansion
$ for different values of $\Delta P_{\mathrm{L}}=P_{\mathrm{L}}-P_{\mathrm{LC}%
}$, where $P_{\mathrm{L}}$ is the laser power and $P_{\mathrm{LC}}$ is the
threshold value. Equation (\ref{P(A_x,A_y)}) together with the measured
probability distribution function $w\left(  X_{\phi}^{\prime}\right)  $ are
employed to evaluate the phase space distribution seen in panel (a) of Fig.
\ref{Fig Wigner vs P_L}. Panel (b) of Fig. \ref{Fig Wigner vs P_L} exhibits
the theoretical prediction for the phase space distribution based on Eq.
(\ref{Wig steady state}). The device parameters that have been employed for
generating the plot in panel (b) are listed in the caption of Fig.
\ref{Fig Wigner vs P_L}.

In another on-fiber optomechanical cavity (sample B) having a FBG mirror
\cite{Zaitsev_046605}, the dependence on laser wavelength was investigated.
The experimental results are compared with theory in Fig.
\ref{Fig Wigner vs WL}. The device parameters that have been employed for
generating the plot in panel (b) are listed in the caption of Fig.
\ref{Fig Wigner vs WL}. In both panels the results are presented as a function
of the detuning factor $s_{\mathrm{D}}\equiv4\pi x_{\mathrm{D}}/\lambda
\beta_{+}$. Note that positive values of $s_{\mathrm{D}}$ correspond to 'red'
detuning, i.e. $\lambda>\lambda_{\mathrm{R}}=1.5451702%
%TCIMACRO{\unit{\U{3bc}m}}%
%BeginExpansion
\operatorname{\mu m}%
%EndExpansion
$, where $\lambda_{\mathrm{R}}$ is the cavity resonance wavelength (see
caption of Fig. \ref{Fig Wigner vs WL}).

In summary, tomography is employed to measure phase space distribution near
the threshold of self-excited oscillation. The comparison with theory allows
the extraction of device parameters, which in turn can be used to evaluate the
expected sensitivity of sensors operating in the region of self-excited oscillation.

This work was supported by the Israel Science Foundation, the bi-national
science foundation, the Deborah Foundation, the Robert J. Shillman Foundation,
the Mitchel Foundation, the Israel Ministry of Science, the Russell Berrie
Nanotechnology Institute, the European STREP QNEMS Project, MAGNET Metro 450
consortium and MAFAT.

%Just because of unusual number of tables stacked at end
\bibliographystyle{apsrmp}
\bibliography{acompat,Eyal_Bib}
%Produces the bibliography via BibTeX.

\end{document}